\documentclass[copyright,creativecommons]{eptcs}
\providecommand{\event}{GandALF 2015}
\usepackage{breakurl}

\usepackage{amsmath}
\usepackage{amsthm}
\usepackage{amsfonts}
\usepackage{stmaryrd}
\usepackage{color}
\usepackage{graphicx}
\usepackage{tikz}
\usetikzlibrary{arrows,positioning,backgrounds,calc,shapes,decorations}
\usetikzlibrary{decorations.pathmorphing}
\usepackage{multirow}
\usepackage{paralist}
\usepackage{caption}
\usepackage{subcaption}

\newcommand{\comment}[1]{}

\newcommand{\mc}[1]{\mathcal{#1}}
\newcommand{\set}[1]{\{#1\}}

\newcommand{\src}{\mathit{src}}
\newcommand{\trg}{\mathit{trg}}

\newcommand{\reach}{\mathsf{Reach}}
\newcommand{\nprop}{\mathsf{NodeProperty}}

\newcommand{\paths}{\mathsf{Paths}}

\newcommand{\calph}{\Sigma_\mathsf{e}}
\newcommand{\galph}{\Sigma_\mathsf{t}}

\newcommand{\last}{\mathsf{last}}

\newtheorem{example}{Example}
\newtheorem{proposition}{Proposition}
\newtheorem{definition}{Definition}
\newtheorem{theorem}{Theorem}

\def\papertitle{Reachability Analysis of Reversal-bounded Automata on Series-Parallel Graphs}

\title{\papertitle}

\author{Rayna Dimitrova
\institute{Max Planck Institute for Software Systems (MPI-SWS), Germany}
\and Rupak Majumdar
\institute{Max Planck Institute for Software Systems (MPI-SWS), Germany}}
\def\authorrunning{R. Dimitrova and R. Majumdar}
\def\titlerunning{\papertitle}

\begin{document}

\maketitle

\begin{abstract}
Extensions to finite-state automata on strings, such as multi-head automata or multi-counter automata,
have been successfully used to encode many infinite-state non-regular verification problems.
In this paper, we consider a generalization of automata-theoretic 
infinite-state verification from strings to labeled \emph{series-parallel graphs}.
We define a model of non-deterministic, 2-way, concurrent automata working on series-parallel graphs and communicating through
shared registers on the nodes of the graph.
We consider the following verification problem: given a family of series-parallel graphs described
by a context-free graph transformation system (GTS), and a concurrent automaton over series-parallel graphs,
is some graph generated by the GTS accepted by the automaton?
The general problem is undecidable already for (one-way) multi-head automata over strings.
We show that a {\em bounded} version, where the automata make a fixed number of reversals along the graph
and use a fixed number of shared registers is decidable, even though there is no bound on the sizes of series-parallel
graphs generated by the GTS.
Our decidability result is based on establishing that the number of context switches is bounded and on an encoding of the computation of bounded concurrent automata to reduce the emptiness problem to the emptiness problem for pushdown automata.
\end{abstract}

\section{Introduction}
The language-theoretic approach to verification 
models the behaviors of a system as a set ---or a \emph{language}--- of structures (such as strings or trees),
and defines machine models that \emph{generate} or \emph{accept} these languages.
The verification problem reduces to the language-emptiness problem for these models.
The simplest such models are finite-state machines over finite or infinite words or trees, and this forms
the basis of the hugely successful {\em automata-theoretic approach} to (finite-state) model checking \cite{Vardi}.
Finite state machines have been generalized in many ways to extend the set of languages that may be needed 
to model more complex (non-regular) computational processes. 
For example, they can be extended with data structures such as stacks or counters, or with multiple heads or tapes and
allowing 2-way traversals of the input \cite{Rosenberg/65/OnMultiHeadFiniteAutomata,RabinS/59/FiniteAutomataAndTheirDecisionProblems},

Since the emptiness problem can be undecidable for many extensions, research in infinite-state verification
has focused on finding suitable underapproximations for which language emptiness is algorithmically
decidable.
For example, the \emph{reversal boundedness} restriction bounds the number of reversals of the counters or of stacks,
or the number of traversals of the input \cite{Ibarra/78/ReversalBoundedMulticounterMachines,Ibarra/14/AutomataWithReversalBoundedCounters,GurariIbarra,GurariI81}
and the \emph{bounded language} restriction considers behaviors describable by a bounded language
\cite{EsparzaGanty,EsparzaGantyMajumdarLICS2012}.
Overall, the approach has led to beautiful theoretical results and has also
been quite successful in modeling many infinite-state parameterized computational models 
and reasoning about them algorithmically. 

Most previous work in parameterized verification
has focused on machine models for string or tree languages.
In this paper, we study behaviors encoded as \emph{series-parallel} graphs whose edges
are labeled with a finite alphabet.
Series-parallel graphs generalize strings or multi-tape machines 
by allowing multiple parallel ``tracks'' to fork off and rejoin at any point.
They allow modeling various natural modes of computation, e.g., fork-join parallelism in
data-parallel programs, while retaining enough structure, e.g., having a natural ``forward''
direction, that is absent in general graphs.
Languages over series-parallel graphs can be naturally described using context-free graph 
transformation systems (GTSs), which describe the dynamic evolution
of families of graphs through local rewrite 
rules 
\cite{CourcelleE/12/GraphStructureAndMSO,BaldanCK/01/StaticAnalysisTechniqueForGTS,BertrandDKSS/12/OnTheDecidabilityStatusOfReachabilityAndCoverabilityInGTS}.

We define and study a class of concurrent finite-state automata traversing series-parallel graphs and communicating
through state-holding registers located at the nodes of the graph.
More precisely, in our model of computation, a fixed number of finite-state machines traverse the nodes of a series-parallel
graph. 
At each step, one of the machines makes a transition that depends on the current state of the machine,
the label it reads on one of the incoming or outgoing arcs, and the value of the register stored at its node.
The machine moves along the selected edge, updating its state as well as the register.
Machines are thus 2-way and non-deterministic, and communicate through the shared registers.
A series-parallel graph is accepted if some subset of machines reaches some final states 
being at the same node of the graph.

We study the emptiness problem: given a context-free GTS defining a language of series-parallel graphs,
and a concurrent finite-state automaton, check if there is a graph in the language of the GTS accepted by the automaton.
This problem is, not surprisingly, undecidable: for example, we can encode linear bounded automata
over strings.
We study a natural restriction of the emptiness problem by restricting the number of reversals along the computation
and by putting a bound on the number of shared registers in the graph.
With these two restrictions, we show that the emptiness problem is decidable and
can be reduced to the emptiness problem for pushdown automata.
Note that even with the restrictions, the problem is infinite-state because there is no a priori bound
on the size of the series-parallel graphs generated by the GTS.

The reduction is based on two technical observations.
First, when the number of reversals and the number of registers are fixed, 
there is a bound on the number of parallel tracks in the graph that needs to be tracked.
We also establish a bound on the number of different times each machine
moves along the run (although the length of the run may be unbounded).
Second, using the bounds above, we construct a large alphabet that tracks valid runs of the
machines on a valid graph generated by the GTS.
We do this in several steps.
We construct a pushdown automaton that checks that a word is a valid representation of a subgraph of a graph
generated by the context-free GTS.
We construct a set of automata, one for each machine, that checks that the word encodes a correct
run of that machine along the graph.
Finally, we construct another automaton that checks that the run is accepted by the concurrent automaton.
Some graph generated by the GTS is accepted if the intersection of all these automata is non-empty.

\paragraph{Other Related Work} 
The automata-theoretic approach is often called {\em regular} model checking, when applied to parameterized verification \cite{SurveyRMC}.
An extensive study of the decidability of several verification problems for classes of GTSs was carried out in~\cite{BertrandDKSS/12/OnTheDecidabilityStatusOfReachabilityAndCoverabilityInGTS}. The problems considered there are \emph{reachability} of a given graph, \emph{coverability} (reachability of a graph that contains a given graph as a subgraph) and \emph{existential coverability}, which asks whether there exists an initial graph such that the answer to the coverability problem is positive. The classes of GTSs they investigate are defined by structural restrictions on the set of transformation rules. Classes with decidable coverability problem are context-free graph grammars, well-structured GTSs and the ones that keep the number of nodes constant. 
\emph{Hyperedge-replacement graph grammars}~\cite{DrewesKH/97/HyperedgeReplacementGraphGrammars} and \emph{vertex-replacement graph grammars}~\cite{EngelfrietR/97/NodeReplacementGraphGrammars} are well-studied classes of GTSs. It is known that for such graph grammars satisfiability of Monadic Second Order (MSO) formulas is decidable~\cite{CourcelleE/12/GraphStructureAndMSO}. 
A logic for expressing properties that involve interleaving of temporal and graph modalities was developed in~\cite{BaldanCKL/06/TemporalGraphLogic} as a combination of MSO and the $\mu$-calculus. They employ an approximation of GTS~\cite{BaldanCK/01/StaticAnalysisTechniqueForGTS} that preserves fragments of the logic to obtain a sound but incomplete verification method for these fragments. A method to refine such approximations based on counterexamples was developed in~\cite{KonigK/06/CEGARGTS}. \cite{Rensink/08/ModelCheckingGraphGrammars} describes a tool for model checking finite-state graph transition systems against first order temporal logic properties.
The work~\cite{MadhusudanP11/auxiliarystorage} studies the emptiness problem for concurrent automata with auxiliary storage
and provides a generalization of the decidability results for a number of classes of such automata
for which the emptiness problem can be reduced to emptiness of finite-state graph automata defined 
MSO definable graphs with bounded tree width.
It might be possible to obtain or generalize the results we establish in this paper through arguments similar to theirs.

\section{Graph-grammar Transition Systems}
Let $A$ be a finite set. As usual, the set $A^*$ consists of all finite sequences of elements of $A$. 
Let $\pi = a_0a_1\ldots a_{n-1}a_n \in A^*$. 
We define $\pi^{-1} = a_na_{n-1} \ldots a_1a_0$ and $\last(\pi) = a_n$. 
The length $|\pi| = n+1$ of $\pi$ is the number of elements of $\pi$ and 
given $0 \leq i \leq j \leq n$ we denote $\pi[i] = a_i$ and $\pi[i,j] = a_i \ldots a_j$.

With $\mathbb{M}(A) = \{S \mid S : A \to \mathbb N\}$ we denote the set of multisets over $A$. 
For $S_1,S_2 \in \mathbb{M}(A)$ we define $S_1 \preceq S_2$ iff for every 
$a \in A$ we have $S_1(a) \leq S_2(a)$. 
We use square brackets to denote multisets, for example, $[a_1,a_2,a_2]$ 
denotes $S \in \mathbb{M}(A)$, where $S(a_1)  = 1$, $S(a_2)  = 2$ and $S(a)  = 0$ for all $a \in A\setminus\{a_1,a_2\}$.

\subsection{Series-parallel graph grammars}

Fix an alphabet $\Sigma$.
We consider graphs labeled with letters from $\Sigma$.
A \emph{graph} is a tuple $G = (N, E, n_b, n_e)$ where 
$N$ is a finite set of nodes,
$E \in \mathbb{M}(N \times N \times \Sigma)$ is a multiset of edges
and $n_b, n_e\in N$ are two distinguished nodes called source and sink, respectively.
For an edge $e = (n, n',\sigma)\in E$, we write $\src(e)$ for $n$ and $\trg(e)$ for $n'$, and 
$\alpha(e)$ for the label $\sigma$ of $e$.
We write $\mc H_{\Sigma}$ for the set of all $\Sigma$-labeled graphs.

Let $G=(N,E,n_b,n_e)$ and $G' = (N', E', n_b', n_e')$ be graphs on disjoint sets of nodes.
For an edge $\widehat e = (\widehat n_1,\widehat n_2,\widehat\sigma) \in E$,
the \emph{edge replacement} graph $G[\widehat e \mapsto G']$ is the (unique up to isomorphism) graph  
defined by removing one copy of the edge $\widehat e$ from $G$, 
and adding the nodes and edges of $G'$ by fusing $\widehat n_1$ with $n_b'$, and $\widehat n_2$ with  $n_e'$. 
Formally, 
$G[\widehat e \mapsto G'] = (N'',E'',n_b,n_e)$, where 
$N'' = N \;\dot\cup\; (N'\setminus \{n_b',n_e'\})$, 
$E'' = (E \setminus \{\widehat e \}) \cup \widehat{E}'$, where
 there is an edge $(n_1, n_2,\sigma)$ in the multiset $\widehat{E}'$ with some multiplicity iff
 \begin{compactitem}
 \item in $E'$, with the same multiplicity, there is an edge $(n_b', n_e',\sigma)$, and 
  $n_1 = \src(\widehat e)$ and $n_2 = \trg(\widehat e)$, or 
 \item in $E'$, with the same multiplicity, there is an edge $(n_1, n_2,\sigma)$, and
   $n_1 \not = n_b'$ and $n_2 \not = n_e'$, or
 \item in $E'$, with the same multiplicity, there is an edge $(n_b', n_2,\sigma)$, and 
  $n_1 = \src(\widehat e)$, or
 \item in $E'$, with the same multiplicity, there is an edge $(n_1, n_e',\sigma)$, and
 $n_2 = \trg(\widehat e)$. 
 \end{compactitem}

\begin{definition}[Series-parallel graph grammar]
A \emph{series parallel graph grammar} (SPGG) is a tuple $\mc G = (V,\Sigma,R,G_0)$, where 
$V$ is a finite set of \emph{variables},
$\Sigma$ is a finite alphabet ($\Sigma \cap V = \emptyset$),
$R \subseteq V \times \mc H_{\Sigma \cup V}$ is a finite set of \emph{rules}, 
$G_0 = (\set{n_b, n_e},\set{(n_b, n_e,v_0)},n_b,n_e) \in \mc H_{V}$, with $n_b \neq n_e$ is the \emph{initial graph}. 

Furthermore, each rule $(v,G') \in R$, where 
$G' = (N',E',n_b',n_e')$,  
satisfies exactly one of the following:
\begin{compactitem}
\item[(1)] $N' = \{n_b',n_e'\}$, $E' = \{(n_b', n_e',\sigma)\}$ and $\sigma \in \Sigma$,
denoted $(v,\sigma) \in R$;
\item[(2)] $N' = \{n_b',n_e',n'\}$ has three nodes, $E' = \{(n_b',n',v_1),(n',n_e',v_2)\}$ and $v_1,v_2 \in V$,
denoted by $(v,v_1\cdot v_2) \in R$ \emph{(series composition)};
\item[(3)] $N' = \{n_b',n_e'\}$ has two nodes, $E' = \{(n_b',n_e',v_1),(n_b',n_e',v_2)\}$ and $v_1,v_2 \in V$,
denoted by $(v,v_1\parallel v_2) \in R$ \emph{(parallel composition)}.
\end{compactitem}
\end{definition}

An SPGG derives a graph in $\mc H_{\Sigma}$ as follows. 
It starts with the graph $G_0$. 
In each step, it picks an arbitrary edge $e$ of the current
graph $G$ that is labeled with a variable $v\in V$, and applies a rule $(v, G')\in R$ to get a new
graph $G'' = G[e \mapsto G']$.
In this case, we write $G\Longrightarrow G''$.
A graph $G \in \mc H_{\Sigma}$ is derived if there is a sequence 
$G_0 \Longrightarrow G_1 \ldots \Longrightarrow G_n = G$ of steps that results in $G$.
Note that every graph thus derived is a series-parallel graph labeled with $\Sigma$,
so an SPGG represents a set of series-parallel graphs labeled with $\Sigma$.
We write $\mc L({\mc G})$
for the set of graphs in $\mc H_{\Sigma}$ derived by ${\mc G}$. 

\begin{example}\label{exmp:graph-grammar}
As an example of an SPGG consider
$\mc G = (V,\Sigma,R,G_0)$ with 
variables $V = \{v_0,v_1,v_a,v_b,v_c\}$,
set of terminal symbols $\Sigma = \{a,b,c\}$,
initial graph $G_0 = (\{n_b,n_e\},\{(n_b,n_e,v_0)\},n_b,n_e)$
and rules

\noindent
$
\begin{array}{lll}
R & =  & \{
(v_0,v_c \cdot v_1), 
(v_1,v_a \parallel v_b),(v_a,a),
(v_b,b),
(v_c,c),
(v_a,v_a \cdot v_a), 
(v_b,v_b \cdot v_b),
(v_b,v_b \parallel v_b),
\}.
\end{array}
$

Figure~\ref{fig:spgraph} shows a (series-parallel) graph $G$ derived from the SPGG $\mc G$. 
The directions of the edges denote the ``natural'' direction from source to sink associated with
a series parallel graph.

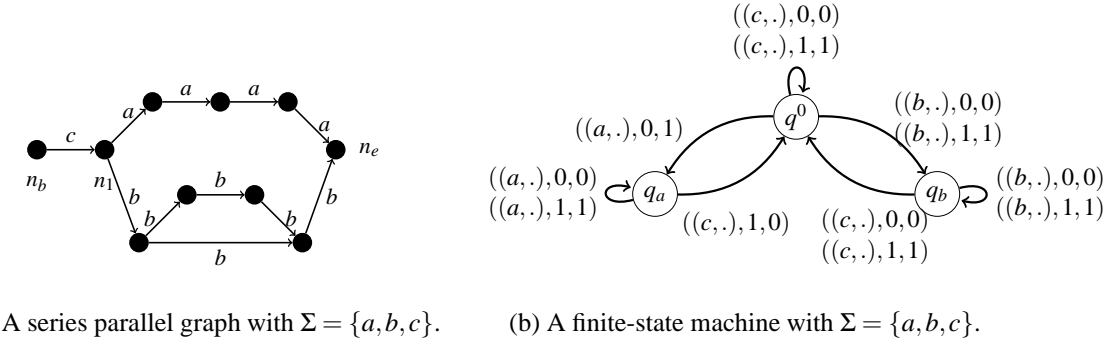
\begin{figure}
\begin{subfigure}[b]{0.45\textwidth}
\begin{center}
\scalebox{0.6}{
\begin{tikzpicture} [inner sep=4pt,node distance=1.5cm]
\tikzstyle{node} = [circle,draw=black,line width = 1pt,fill=black]
\tikzstyle{reg} = [draw=black,rectangle, inner sep=1.7pt,line width = 1pt]

\tikzstyle{state} = [circle,draw=black,line width = 2pt,scale=0.5]

\node[node] (nb) at (0,0) {}; 
\node[node, right of=nb] (n1) {};
\node[node, above right of=n1,xshift=.0cm] (s1) {};
\node[node, below right of=n1,yshift=-1cm,xshift=-.3cm] (s2) {};
\node[node, right of=s1,xshift=.0cm] (s3) {};
\node[node, right of=s3,xshift=.0cm] (s4) {};
\node[node, below right of=s4,xshift=.0cm] (ne) {};
\node[node, above right of=s2,xshift=0cm] (s5) {};
\node[node, right of=s5,xshift=0cm] (s6) {};
\node[node, below right of=s6,xshift=0cm] (s7) {};

\path[draw = black,line width=1pt] 
(nb) edge[->] node[above,midway]{\Large $c$}(n1)
(n1) edge[->] node[above,midway]{\Large $a$}(s1)
(n1) edge[->] node[right,midway]{\Large $b$} (s2)
(s1) edge[->] node[above,midway]{\Large $a$} (s3)
(s3) edge[->] node[above,midway]{\Large $a$} (s4)
(s4) edge[->] node[right,midway]{\Large $a$} (ne)
(s2) edge[->] node[left,midway]{\Large $b$} (s5)
(s5) edge[->] node[above,midway]{\Large $b$} (s6)
(s6) edge[->] node[right,midway]{\Large $b$} (s7)
(s2) edge[->] node[below,midway]{\Large $b$} (s7)
(s7) edge[->] node[right,midway]{\Large $b$} (ne)
;

\node[below of=nb,yshift=.75cm] {\Large$n_b$};
\node[below of=n1,yshift=.75cm] {\Large$n_1$};
\node[right of=ne,xshift=-.75cm] {\Large$n_e$};

\end{tikzpicture}
}
\end{center}
\caption{A series parallel graph with $\Sigma = \{a,b,c\}$.}
\label{fig:spgraph}
\end{subfigure}%
\begin{subfigure}[b]{0.45\textwidth}
\begin{center}
\scalebox{0.85}{
\begin{tikzpicture} 
\tikzstyle{state} = [circle,draw=black,inner sep=.5pt, minimum size=.7cm]

\node[state] (q0) at (0,0) {$q^0$}; 
\node[state, below left of=q0,xshift=-1.5cm,yshift=-.5cm] (qa) {$q_a$};
\node[state, below right of=q0,xshift=1.5cm,yshift=-.5cm] (qb) {$q_b$};

\path[draw = black,line width=1pt] 
(q0) edge[->,bend right] node[left,midway,text width=2cm]{$((a,.),0,1)$}(qa)
(q0) edge[->,bend left] node[right,near start,text width=2cm,xshift=.5cm]{$((b,.),0,0)$\\$((b,.),1,1)$}(qb)
(qa) edge[->,bend right] node[right,near start,xshift=-.6cm,yshift=-.5cm,text width=2cm]{$((c,.),1,0)$}(q0)
(qb) edge[->,bend left] node[left,near start,xshift=1.2cm,yshift=-.7cm,text width=2cm]{$((c,.),0,0)$\\$((c,.),1,1)$}(q0)
(q0) edge[->,loop above] node[above
,midway,text width=2cm]{$((c,.),0,0)$\\$((c,.),1,1)$}(q0)
(qa) edge[->,loop left] node[left,midway,text width=2cm,xshift=.3cm]{$((a,.),0,0)$\\$((a,.),1,1)$}(qa)
(qb) edge[->,loop right] node[right,midway,text width=2cm]{$((b,.),0,0)$\\$((b,.),1,1)$}(qb)
;
\end{tikzpicture}
}
\end{center}
\caption{A finite-state machine with $\Sigma = \{a,b,c\}$.}
\label{fig:fsm}
\end{subfigure}%
\caption{A series parallel graph generated by an SPGG and a finite-state machine over the same alphabet.}
\end{figure}
\end{example}

A series-parallel graph has a natural ``direction'' associated with it from the source to the sink, consistent
with the direction $n_1 \rightarrow n_2$ of an edge $(n_1, n_2,\sigma)$.
In particular, it has no directed cycles.
For convenience, we introduce the ``symmetric closure'' of series-parallel graphs.
For each edge $(n_1, n_2,\sigma)$ labeled with $\sigma$, we augment the label with a direction $1$ 
to obtain $(\sigma,1)$ ($1$ capturing the ``forward'' direction),
and add an opposite edge $(n_2, n_1,(\sigma,-1))$ labeled with $(\sigma, -1)$ denoting the edge taken in the ``backward'' direction.
Formally, given a series-parallel graph $G = (N, E, n_b, n_e)$, we define its symmetric closure
$G' = (N, E', n_b, n_e) \in \mc H_{\Sigma\times\set{1,-1}}$, where
$E' = \set{(n, n',(\sigma,1)) \mid (n, n',\sigma) \in E} \cup
\set{(n, n',(\sigma,-1)) \mid (n', n,\sigma) \in E}$.
We write $\mc L^u({\mc G})$ for the set of symmetric closures of all graphs derived by ${\mc G}$.

\smallskip
\noindent
\textit{Remark.}
While for simplicity of the presentation we consider graphs with a single pair of source and sink nodes,
our results can in principle be extended to graphs with multiple such nodes. However, the automata construction
outlined in Section~\ref{sec:pda-traces} relies on the structure of the rules of an SPGG and does not directly 
generalize to general context-free GTSs defining sets of directed acyclic graphs.  
\subsection{Graph-grammar transition systems}

We now define communicating finite automata on the symmetric closure of
series-parallel graphs. Recall that these are series-parallel graphs whose
edges are labeled with an alphabet and a direction.
Intuitively, a system of communicating machines has a set of $m$ machines
that traverse the edges of a series-parallel graph, some of whose nodes are annotated with
Boolean registers.
Each automaton traverses the edges of the graph:
when the automaton is at a node $n$ of the graph and in state $q$, it 
reads the register on the node,
chooses an edge with source node $n$ labeled with $(\sigma,d)\in \Sigma \times \{1,-1\}$ based
on its current state, the label, and the value read from the register,
traverses the edge and moves to the target node of that edge and to a new state $q'$, and 
writes a value to the register at the source node.

Let $\Sigma$ be a finite alphabet.
A \emph{finite-state machine} $\mc M = (Q,q^0,\Sigma,\delta)$ 
consists of  
finite set of states $Q$,
initial state $q^0 \in Q$,
input alphabet $\Sigma$, and
transition relation
$\delta \subseteq 
Q \times (\Sigma\times \{1,-1\}) 
  \times \mathbb{B} \times Q \times \mathbb{B}$.
 
The intuitive meaning of a transition $(q,(\sigma,d),b,q',b') \in \delta$ is that when
the machine $\mc M$ is in state $q$ and reads input letter $\sigma \in \Sigma$, direction $\set{1,-1}$,
and register value $b$, then it
changes its state to $q'$ and moves along an edge labeled $(\sigma, d)$ in the graph and writes $b'$
to the register.

\begin{example}\label{exmp:fsm}
Figure~\ref{fig:fsm} shows an example of a finite-state machine $\mc M = (Q,q^0,\Sigma,\delta)$
with states $Q = \{q^0,q_a,q_b\}$, input alphabet $\Sigma = \{a,b,c\}$ and transition relation
$\delta$ depicted in Figure~\ref{fig:fsm}, where a label $(((\sigma,d),p,p')$ on an edge
from state $q$ to state $q'$ stands for the transition $((q,(\sigma,d),p,q',p')$.
\end{example}

A system of machines $(\mc M, m)$ is a set of $m$ disjoint copies $\mc M_1,\ldots, \mc M_m$
of the machine $\mc M$.

\begin{definition}[Graph-grammar transition system]\label{def:transition-system}
Let $\mc M = (Q, q^0, \Sigma, \delta)$ be a finite-state machine. 
A system of $m$ machines $(\mc M, m)$, together with an 
SPGG $\mc G = (V,\Sigma,R,G_0)$ 
defines a \emph{transition system} $T({\mc M}, m, {\mc G}) = (\Gamma,\Gamma_0,\rightarrow)$ as follows.
The set of \emph{configurations} $\Gamma$ consists of all tuples 
$\langle G,\mu, \beta \rangle$ such that $G \in \mc L^u({\mc G})$ is a graph derived by $\mc G$ and:
\begin{compactitem}

\item $\mu : N \to 2^{\{1,\ldots,m\} \times Q}$ maps each node in $G$ to the states of the machines at that node;
we require that for each $i \in \{1,\ldots,m\}$ there exists exactly one $n \in N$ and exactly one $q \in Q$ with 
$(i,q) \in \mu(n)$;
\item $\beta : N \rightarrow \mathbb{B}$ maps each node to the value of the Boolean register at that node.
\end{compactitem}

The set $\Gamma_0$ of \emph{initial configurations} is such that $\gamma = \langle G,\mu, \beta \rangle \in \Gamma_0$ iff $\gamma \in \Gamma$, 
$\mu(n_b)  = \{(i,q^0)\mid i \in \{1,\ldots,m\} \}$, 
$\mu(n) = \emptyset$ for every $n \in N \setminus \{n_b\}$, and
$\beta(n) = 0$ for every $n\in N$.
That is, initially all machines are positioned at the source node of the graph and are in their initial state, and all registers are $0$.

The \emph{successor relation} $\rightarrow \subseteq \Gamma \times \Gamma$ is defined as 
$\rightarrow = \bigcup_{i =  1}^{m}\rightarrow_{i}$, where for each $i \in \{1,\ldots,m\}$ it holds that 
$(\langle G,\mu,\beta\rangle, \langle G',\mu',\beta'\rangle) \in \rightarrow_i$, (denoted  $\langle G,\mu,\beta\rangle \rightarrow_i \langle G',\mu',\beta'\rangle$) 
iff the following hold:
\begin{compactitem}
\item $G' = G$, where $G = (N,E,n_b,n_e) \in {\mc L}^u(\mc G)$ is a graph generated by $\mc G$. 
\item There exist an edge $e = (n,n',(\sigma,d)) \in E$, 
states $q, q' \in Q$, and a value $b'\in {\mathbb B}$ such that:
\begin{compactitem}
\item[(i)]
$(i,q) \in \mu(n)$ (Note: $n \neq n'$, since $\mc G$ is an SPGG),
\item[(ii)] 
$(q,\alpha(e),\beta(n),q',b') \in \delta$, 
\item[(iii)]
$\mu'(n) = \mu(n)\setminus \{(i,q)\}$, $\mu'(n') = \mu(n)\cup \{(i,q')\}$ and
$\mu'(n'') = \mu(n'')$ for all $n'' \in  N \setminus \{n,n'\}$,
\item[(iv)]
$\beta'(n) = b'$ and $\beta'(n'') = \beta(n'')$ for all $n'' \in N\setminus\set{n}$.
\end{compactitem} 
We say that the edge $e$ is \emph{compatible} with the transition $\gamma \rightarrow \gamma'$.
\end{compactitem}

\end{definition}

A \emph{run} $\rho$ of $T({\mc M},m, {\mc G}) = (\Gamma,\Gamma_0,\rightarrow)$ 
is a sequence of configurations 
$\rho = \gamma_0\ldots\gamma_f \in \Gamma^*$ such that $\gamma_0 \in \Gamma_0$ and $\gamma_{i-1} \rightarrow \gamma_{i}$ for each $i = 1,\ldots,f$.

Intuitively, the infinite-state transition system $T({\mc M}, m, {\mc G})$ captures the behaviors of $m$ machines, copies of $\mc M$, on the family
of all series-parallel graphs derived by ${\mc G}$.

\subsection{Configuration properties and verification problem} 
A {\em configuration property} describes a set of configurations. 
Let $n$ be a variable (ranging over nodes), 
and $S \in \mathbb{M}(Q)$.
The set of \emph{configuration properties} consists of the positive Boolean combinations (no negation)
of {\em atomic} properties of the form $\exists n.\ S \preceq \mu(n)$.

A configuration $\gamma = \langle G,\mu,\beta\rangle \in \Gamma$ with 
$G = (N,E,n_b,n_e)$ satisfies 
an atomic configuration property $\varphi = \exists n.\ S \preceq \mu(n)$ (written $\gamma \models \varphi$) iff there exists a node $n \in N$ such that $S \preceq [q \in Q \mid (i,q) \in \mu(n) \text{, where } i \in \{1,\ldots,m\}]$, that is, the multiset $S$ is contained in the multiset of machine states in the node $n$ in the configuration $\gamma$. 
The relation $\models$ is naturally extended to positive Boolean combinations. 

Let $({\mc M},m)$ be a system of machines and $\mc G$ an SPGG.
Given a configuration property 
$F$ describing a set of \emph{final configurations},
the \emph{verification problem}
$\reach({\mc M},m, {\mc G}, F)$
is to decide whether there exists a run 
$\rho = \gamma_0 \ldots \gamma_f$ of $T({\mc M}, m, {\mc G})$ 
such that $\gamma_i \models F$ for some $0 \leq i \leq f$, i.e., a run that reaches $F$.

Since our model allows machines to do arbitrarily many ``reversals'' (i.e., following
forward and backward edges) and do not fix a bound on the number of shared registers that are read or written, 
it easily captures linear bounded automata. Thus, the verification problem is in general undecidable.

\begin{proposition}
The verification problem is undecidable.
\end{proposition}

\subsection{Bounded verification problem}\label{sec:bounded-verification} 

Since the general problem is undecidable, we focus on a bounded version.
We introduce two restrictions.
First, we allow each machine to make only a bounded number of reversals (a reversal occurs when the machine
changes direction in the graph).
Second, we fix an a priori bound on the number of shared registers. That is, while the SPGG generates a potentially unbounded
set of graphs, with unboundedly many nodes, we assume that there is some fixed bound $k$ on the number of Boolean registers located at nodes of a generated graph
(these $k$ registers may be situated at arbitrary nodes of the graph though).

Fix a machine ${\mc M} = (Q, q^0, \Sigma, \delta)$,
the system of machines $({\mc M},m)$, and an SPGG ${\mc G} = (V,\Sigma, R, G_0)$.

\paragraph{Reversal bound.}
Let us fix a run $\rho = \gamma_0,\ldots,\gamma_f$ where
$\gamma_i = \langle G,\mu_i,\beta_i\rangle$.
Consider the projection of $\rho$ to $\rightarrow_j$ for each machine $j\in \set{1,\ldots, m}$.
The number of {\em reversals} made by machine $j$ along the run, intuitively, is 
the number of times it changes from traversing an edge marked with direction $1$ to traversing
an edge marked with direction $-1$, or vice versa.

Formally,
let $e_1e_2\ldots e_n$ be a sequence of edges.
A reversal occurs at position $i$ if 
$\alpha(e_i) = (\cdot, 1)$ and $\alpha(e_{i+1}) = (\cdot, -1)$
or if 
$\alpha(e_i) = (\cdot, -1)$ and $\alpha(e_{i+1}) = (\cdot, 1)$.

Now, let $\gamma_{i_1} \rightarrow_j \gamma_{i_1 + 1}$, $\gamma_{i_2} \rightarrow_j \gamma_{i_2 + 1}$, $\ldots$
be the transitions of machine $j$ along the run $\rho$, and let $e_{i_1}$, $e_{i_2}$, $\ldots$ be the
compatible edges that were taken by machine $j$.
The number of reversals of machine $j$ along $\rho$ is the number of reversals in the sequence $e_{i_1} e_{i_2}\ldots$.

For $r \geq 0$, the set of \emph{$r$-reversal bounded} runs of $T({\mc M}, m, {\mc G})$ is the set of runs
in which each machine makes at most $r$ reversals.

\paragraph{Register bound.}
The register bound fixes a number $k$ of Boolean registers.
That is, each graph $G$ derived by $\mc G$ comes with a mapping $\kappa: N \rightarrow \set{0,1}$,
such that $|\kappa^{-1}(1)| \leq k$, and we allow the machines to read and write register values only
when their current node is in $\kappa^{-1}(1)$.

To derive graphs with a mapping $\kappa$,
we modify an SPGG to ``mark'' some nodes along the derivation, and ensure that any derived graph 
has at most $k$ marked nodes. (The formal details are similar to constructing a CFG for a CFL 
with at most $k$ marked positions from a CFG for the (unmarked) language.)
For an SPGG $\mc G$, we denote by ${\mc G}^k$ the SPGG that marks at most $k$ nodes of a derived graph.
We write, by abuse of notation, $(G,\kappa) \in \mc L^u(\mc G^k)$ for a graph $G$ which is the
symmetric closure of a graph derived by $\mc G^k$ together with the mapping $\kappa$.

In addition, we modify the successor relation of the graph-grammar transition systems $T(\mc M,m,{\mc G}^k)$ to require (ii)' $(q,\alpha(e), \beta(n), q', b')\in\delta$ if $\kappa(n)= 1$
and $(q, \alpha(e), 0, q', 0) \in \delta$ otherwise.

\begin{example}
The SPGG $\mc G$ shown in Example~\ref{exmp:graph-grammar} can be modified into an SPGG $\mc G^2$ that derives graphs in 
which at most 2 nodes are marked.
Furthermore, we can consider SPGGs that not only ensure an upper bound on the number of marked nodes, but impose constraints on their location. For example, we can consider an SPGG $\mc G^2_*$ that additionally requires that by applying the rule 
$(v_0,v_c \cdot v_1)$ the corresponding intermediate node between the edges labeled $v_c$ and $v_1$ is marked to contain a register.

If we then consider the graph-grammar transition system $T(\mc M,2,{\mc G}^2_*)$, where $\mc M$ is the finite-state machine described in Example~\ref{exmp:fsm}, and let $G$ be the graph depicted in Figure~\ref{fig:spgraph}, then 
there does not exist a run with underlying graph $G$ that reaches a configuration satisfying
$\varphi = \exists n.\ [q_a,q_a] \preceq \mu(n)$, since the register at node $n_1$ acts as a semaphore
that does not allow two copies of the machine $\mc M$ to enter the part of the graph containing edges labeled with the letter $a$.
\end{example} 

\smallskip
The reversal-bounded and register-bounded verification problem takes as input a system of machines $({\mc M}, m)$,
an SPGG $\mc G$, and parameters $r$ and $k$, and a configuration property $F$, and 
asks if there exists an $r$-reversal bounded run of the machines on some graph derived by ${\mc G}^k$
that reaches $F$.

Our main result is the following.

\begin{theorem}
The reversal- and register-bounded verification problem is decidable.
\end{theorem}

\smallskip
\noindent
\textit{Remark.}
Our decidability results hold for a somewhat more general model, in which the machines can read one of the fixed
number of registers not at its current node but can only write to the register at its current node, or vice versa.
We work in the simpler setting to keep the notation manageable.

\section{Properties of Reversal-Bounded Runs}\label{sec:runs-to-words}
Fix a machine ${\mc M} = (Q, q^0, \Sigma, \delta)$,
the system of machines $({\mc M},m)$, an SPGG ${\mc G} = (V,\Sigma, R, G_0)$
and the parameters $r$ and $k$.
In this section we state two properties of $r$-reversal bounded runs of $T({\mc M}, m, {\mc G}^k)$ that allow us to encode such runs as words over a finite alphabet and to reduce the reversal- and register-bounded verification problem to the emptiness test for a context free language.

Given a run $\rho=\gamma_0\ldots\gamma_f$ and a machine $i \in \{1,\ldots,m\}$, an \emph{$i$-block} is 
a segment $\rho[j_1,j_2] = \gamma_{j_1}\ldots\gamma_{j_2}$ of the run $\rho$  such that 
$\gamma_j \rightarrow_i \gamma_{j+1}$ for each $j_1 \leq j < j_2$. 
That is, all transitions in the part $\rho[j_1,j_2]$ of the run are made by machine $i$.
The following proposition establishes that for every $r$-reversal bounded run $\rho$ we can reorder its transitions to obtain an $r$-reversal bounded run $\widehat \rho$ such that the number of maximal blocks in $\widehat \rho$ is not greater than a constant depending on $m,r$ and $k$ (and not on the length of the run $\rho$). 

\begin{proposition}\label{prop:bound-p}
For every $r$-reversal bounded run $\rho = \gamma_0,\ldots,\gamma_f$ of $T({\mc M}, m, {\mc G}^k)$
there exist an $r$-reversal bounded run $\widehat \rho = \widehat\gamma_0,\ldots,\widehat\gamma_f$ of $T({\mc M}, m, {\mc G}^k)$ and
a sequence of indices $0  = f_0 < f_1 < \ldots < f_u = f$ such that the following conditions are satisfied:
\begin{compactitem}
\item $u \leq \big(r \cdot m + k \cdot m \cdot (r+1)+1\big) \cdot (m + 1)$,
\item for each $i \in \{0,\ldots,u-1\}$, there exists $m_i \in \{1,\ldots,m\}$ such that $\widehat\rho[f_i,f_{i+1}]$ is a $m_i$-block, 
\item $\widehat\gamma_0 = \gamma_0$ and $\widehat\mu_f = \mu_f$, where  
$\gamma_f =  \langle G, \mu_f,\beta_f \rangle$ and 
$\widehat\gamma_f =  \langle G, \widehat\mu_f,\widehat\beta_f \rangle$.
\end{compactitem}
\end{proposition}

\smallskip
\noindent
\textit{Remark.}
In the proof of the above proposition we construct the run $\widehat \rho$ by reordering transitions in $\rho$ while keeping in place the transitions that access registers. Thus, the relative order of transitions which modify registers is preserved, which in turn implies that $\widehat\gamma_f = \gamma_f$ (that is, we have also $\widehat\beta_f = \beta_f$).

The second property uses the bound $r$ on the number of reversals of each machine in an $r$-reversal bounded run $\rho$ to relate $\rho$ to the set of paths in the underlying graph traversed by the machines in $\rho$.

A \emph{trace} $\tau$ is an element of the set $\Sigma^*$.  
A trace $\tau = \sigma_1\ldots\sigma_f$ is \emph{compatible} with a run $\rho = \gamma_0,\ldots,\gamma_f$ if 
there exists a sequence of edges $e_1e_2 \ldots e_f$ compatible with $\rho$ such that 
$\alpha(e_i) = (\sigma_i,\cdot)$ for every $0 < i \leq f$.

Given a graph $G = (N,E,n_b,n_e) \in \mc L^u(\mc G^k)$ and a trace $\tau$
we define $\paths(G,\tau)$ to be the (possibly empty) set of paths from $n_b$ to $n_e$ whose sequence of edge labels is $\tau = \sigma_1\ldots\sigma_f$. Formally, for a sequence of nodes $\pi = n_0 n_1\ldots n_f \in N^*$ we have $\pi \in \paths(G,\tau)$ iff $n_0 = n_b$, $n_f = n_e$ and $(n_{i-1},n_{i},(\sigma_i,1)) \in E$.

Below we establish a property of an $r$-reversal bounded run $\rho = \gamma_0\ldots\gamma_f$
of $T({\mc M}, m, {\mc G}^k)$ and a trace $\tau$ that is compatible with $\rho$. Namely, for each machine $i \in \{1,\ldots,m\}$ the corresponding subsequence $\tau_i$ of $\tau$ can be split into at most $r+1$ segments,
such that each of those segments can be embedded in a trace labelling a simple path from $n_b$ to $n_e$ or from $n_e$ to $n_b$. 
This is formalized in the following proposition, which easily follows from the properties of series-parallel graphs.

\begin{proposition}\label{prop:trace-embed}
Let $\rho$ be an $r$-reversal bounded run  of $T({\mc M}, m, {\mc G}^k)$
and $\tau$ be a trace that is compatible with $\rho$.
Let $\pi_i$ be the sequence of nodes visited in $\rho$ by machine $i \in \{1,\ldots,m\}$, in the order they occur in $\rho$, let $\tau_i$ be the corresponding subsequence of $\tau$, and
$r_i \leq r$ be the number of reversals of machine $i$ in $\rho$.

Then, for each $i \in \{1,\ldots,m\}$ and each $h \in \{1,\ldots,r+1\}$ there exist
traces $\tau_{i,h},\tau'_{i,h},\tau''_{i,h},\tau'''_{i,h} \in \Sigma^*$ and
sequences of nodes $\pi_{i,h}, \pi'_{i,h},\pi''_{i,h},\pi'''_{i,h} \in N^*$
such that the following conditions are satisfied:

\begin{compactitem}
\item $\pi_{i,h} \in \paths(G,\tau_{i,h})$, and
$\tau_{i,h} = \tau'_{i,h}\cdot\tau''_{i,h}\cdot\tau'''_{i,h}$, and
$\pi_{i,h} = \pi'_{i,h}\cdot\pi''_{i,h}\cdot\pi'''_{i,h}$;
\item For each $i \in \{1,\ldots,m\}$ there exist indices 
$0  = j_0 < j_1 < \ldots < j_{r_i+1} = |\pi_i|-1$ such that:
\begin{compactitem}
\item if $1 \leq h \leq r_i+1$ and $h$ is odd, then 
$\tau_i[j_{h-1}+1,j_{h}] = \tau''_{i,h}$ and 
$\pi_i[j_{h-1},j_{h}] = \pi''_{i,h}$;
\item if $1 \leq h \leq r_i+1$ and $h$ is even, then
$\tau_i[j_{h-1}+1,j_{h}] = {\tau''_{i,h}}^{-1}$
$\pi_i[j_{h-1},j_{h}] = {\pi''_{i,h}}^{-1}$.
\end{compactitem}
\end{compactitem}
\end{proposition}

Proposition~\ref{prop:bound-p} allows us to restrict our reasoning to $r$-reversal bounded runs
with at most $\big(r \cdot m + k \cdot m \cdot (r+1)+1\big) \cdot (m+1)$ blocks. 
Proposition~\ref{prop:trace-embed} allows us to reduce from reasoning about graphs derived by $\mc G^k$ 
to reasoning about $r+1$-tuples of traces in such graphs.
Based on these results, we define the two parameters
$p = \big(r\cdot m + k \cdot m \cdot (r+1)+1\big) \cdot (m+1)$ and
$t = \widetilde r \cdot m, \textrm{ where } \widetilde r = r + 1$.

\section{Automata-theoretic Algorithm}\label{sec:algorithm}
In this section we present an automata-theoretic algorithm for solving the reversal- and register-bounded verification problem. Before we give an overview of our algorithm and outline the automata constructions it comprises, we recall some basic definitions from automata theory.

\subsection{Preliminaries}
 A 2-way nondeterministic finite automaton (2NFA) is a tuple
$\mc A = (Q,q^0,\Sigma,\delta,A)$, where 
$Q$ is a finite set of states,
$q^0 \in Q$ is the initial state,
$\Sigma$ is a finite alphabet,
$\delta \subseteq Q \times \Sigma \times  Q \times \{-1,1\}$ is the transition relation and 
$A \subseteq Q$ is a set of accepting states.
$\mc A$ is deterministic iff $\delta$ is a function from $Q \times \Sigma$ to $Q \times \{-1,1\}$.
$\mc A$ is a 1-way NFA (NFA) iff $d = 1$ for each $(q,\sigma,q',d) \in \delta$. 

For $q,q' \in Q$, $w',w'',w''',w'''' \in \Sigma^*$, $\sigma \in \Sigma$ and $\sigma' \in \Sigma \cup \{\epsilon\}$, let $\langle q,w',\sigma,w''\rangle \Rightarrow_{\mc A} \langle q',w''',\sigma',w''''\rangle$ iff $(q,\sigma,q',d) \in \delta$ and
(1) if $d = 1$, then $w''' = w'.\sigma$, $w'' =\sigma'.w''''$, either $\sigma' \in \Sigma$ or $\sigma' = \epsilon$ and $w'''' =\epsilon$ and 
(2) if $d = -1$, then $w'''' = \sigma . w''$, $w' =w'''.\sigma'$, either $\sigma' \in \Sigma$ or $\sigma' = \epsilon$ and $w''' =\epsilon$.

If $\mc A$ is an NFA, we define $\delta(q,w)$ for $w \in \Sigma^*$ in the obvious way.

Let $\vdash \in \Sigma$ and $\dashv \in \Sigma$, where $\vdash \neq \dashv$, be designated symbols and $w \in (\Sigma \setminus \{\vdash,\dashv\})^*$.

If $\mc A$ is a 2NFA, then $w \in \mc L(\mc A)$ iff
$\langle q^0,\epsilon,\vdash, w \dashv \rangle \Rightarrow_{\mc A}^* \langle q,\vdash w \dashv,\epsilon,\epsilon \rangle$ or 
$\langle q^0,\epsilon,\vdash, w \dashv \rangle \Rightarrow_{\mc A}^* \langle q,\epsilon,\epsilon,\vdash w \dashv \rangle$ for some $q \in A$.
If $\mc A$ is an NFA, then $w \in \mc L(\mc A)$ iff $\delta(q^0,\vdash w \dashv) \cap A \neq \emptyset$.

A push-down automaton (PDA) is a tuple $\mc P = (Q,q^0,\Sigma,\Delta,\bot,\delta)$, where
$Q$ is a finite set of states,
$q^0 \in Q$ is the initial state,
$\Sigma$ is a finite input alphabet,
$\Delta$ is a finite stack alphabet,
$\bot$ is the start symbol and
$\delta \subseteq Q \times (\Sigma \cup \{\epsilon\}) \times \Delta \times Q \times \Delta^*$ is the transition relation. For $q,q' \in Q$, $\sigma \in \Sigma \cup \{\epsilon\}$, $w \in \Sigma^*$, $a \in \Delta$, $\alpha,\beta \in \Delta^*$ we define 
$\langle q, \sigma.w,a.\alpha\rangle \Rightarrow_{\mc P} \langle q', w,\beta.\alpha\rangle$ iff
$(q,\sigma,a,q',\beta) \in \delta$.

For a PDA $\mc P$, $w \in \mc L(\mc P)$ iff $\langle q^0,\vdash w \dashv,\bot\rangle \Rightarrow_{\mc P}^* \langle q,\epsilon,\bot \rangle$.

\subsection{Overview of the algorithm}
We now outline the construction of a PDA $\mc A$, which we use in order to reduce the reversal- and register-bounded verification problem to checking emptiness of a PDA. We begin by describing the input of the  automata involved in the construction and then proceed to give an overview of the construction followed by a  
formal definition of the input alphabets of these automata.

\looseness=-1
The automaton $\mc A$ reads words that consist of traces in $\Sigma^*$. 
In order to reflect sufficient information about the corresponding nodes and
registers in the underlying graph, these traces are annotated as follows. 
First, since graphs derived by ${\mc G}^k$ contain at most $k$ registers,
we assume these registers to have unique identifiers from the set $\{1,\ldots,k\}$.
Thus, a triple $(\sigma,j_1,j_2) \in \Sigma\times\{0,\ldots,k\}\times\{0,\ldots,k\}$ 
consists of an edge label $\sigma$ and the identifiers of the registers at the source and target node 
of the edge, where $0$ indicates no register at the respective node.
We add additional annotation to reflect which nodes are shared in the corresponding
paths, that is, positions where paths in the series-parallel graph branch off or join. 

The automaton $\mc A$ reads such annotated traces and checks the existence of a run
by emulating the behaviour of the machines on these traces by guessing an execution
for each of them. 
An \emph{execution} of $\mc M = (Q,q^0,\Sigma,\delta)$ is a sequence  
$\xi = q_0,(\sigma_1,b_1,b_1'),q_1,\ldots,(\sigma_f,b_f,b_f'),q_f$ 
such that $(q_{l-1},\sigma_l,b_l,q_l,d_l,b_l') \in \delta$
for some $d_l \in \{1,-1\}$.
In addition to verifying that each guess is indeed an execution,
$\mc A$ needs to also check that the values written to and read from
the shared registers by different machines are consistent. 

Formally, an annotated trace and executions of the machines  define a
\emph{read-write} sequence $\eta = (j_1,b_1,b'_1),\ldots,(j_f,b_f,b'_f) \in 
(\{0,\ldots,k\} \times \mathbb{B} \times \mathbb{B})^*$, 
where, intuitively, $j_i$ is the location that is read and/or written.
Such a read-write sequence $\eta$ is \emph{valid w.r.t.\ an initial register valuation 
$\beta_0 : \{1,\ldots,k\} \to \mathbb{B}$} iff each read operation reads the value
written by the most recent write operation, or the initial value from $\beta_0$ if it is not overwritten,
that is, for $i \in \{1,\ldots,f\}$ with $j_i > 0$ it holds that 
if there is $i' < i$ such that $j_{i'} =  j_i$, then $b_i = b_{i'}'$
for the largest such $i'$, and otherwise $b_i = \beta_0(j_i)$.

\looseness=-1
Thus, the automaton $\mc A$ accepts tuples of traces in some graph derived by ${\mc G}^k$, annotated with information about registers and about nodes shared by the corresponding paths in the graph. 
$\mc A$ also guesses an execution for each of the $m$ machines. 
The PDA $\mc A$ is constructed as the intersection of a PDA $\mc P_{\mathsf t}$ and an NFA $\mc A_{\mathsf e}$ (Section~\ref{sec:pda-traces}). 
$\mc P_{\mathsf t}$ checks that its input word encodes a tuple of traces in some graph derived by ${\mc G}^k$ and that these are correctly annotated with information about registers and the nodes that are shared among the paths corresponding to these traces. 
The NFA $\mc A_{\mathsf{e}}$ guesses and verifies the executions of the machines. It is obtained as the intersection of $m+2$ NFAs:
$m$ NFAs $A_i$, one for each $i \in \{1,\ldots,m\}$, an NFA $\mc A_{\mathsf c}$ and an NFA $\mc A_{\mathsf s}$.
The NFA $\mc A_i$ verifies that the guess of an execution of machine $i \in \{1,\ldots,m\}$ is correct.
We describe the construction of $\mc A_i$ as a 2NFA (Section~\ref{sec:2nfa-executions}) which is then converted to an NFA using standard techniques~\cite{HopcroftUllman}.
Automaton $\mc A_{\mathsf c}$ checks the validity of the read-write sequence corresponding to the annotated traces and the guessed executions  (Section~\ref{sec:2nfa-read-write}). 
Automaton $\mc A_{\mathsf s}$ (Section~\ref{sec:1nfa-properties}) checks that a configuration in $F$ is reached.
The reversal- and register-bounded verification problem thus reduces to checking emptiness of the language of the constructed automaton $\mc A$.

\looseness=-1
According to Section~\ref{sec:runs-to-words}, it suffices to reason about $t = m \cdot (r+1)$ traces in graphs derived by ${\mc G}^k$.  To this end, we define the \emph{trace alphabet}
$\galph = 
\Big(\big(\Sigma\ \dot{\cup}\ \{\flat\}\big) \times \{0,\ldots,k\}^2 \times
\{1,\ldots,t\} \Big)^t\ \dot{\cup} \ \{1,\ldots,t\}^t.$
Words over $\galph$ are tuples of $t$ traces in some graph $G$, annotated with additional information. Each letter in $\galph$ contains one row for each of the $m$ machines and each of the $\widetilde r = r+1$ paths corresponding to it. There are two types of letters. Each row in a letter of the first type consists of a letter in $\Sigma$ (or the special symbol $\flat$) together with two \emph{register identifiers} in  $\{0,\ldots,k\}$ and a \emph{path index} in $\{1,\ldots,t\}$. The letters of the second type are $t$-tuples of path indices in $\{1,\ldots,t\}$, where equal indices indicate paths sharing a node. 

The \emph{execution alphabet}
$\calph = \big(
\{0,\ldots,p\}\times {\mathbb B} \times {\mathbb B}
 \times
Q \times \{1,\ldots,t\}
\big)^t$
is used to describe tuples of executions, one for each of the $m$ machines. 
Each letter contains $\widetilde r = r+1$ rows for each machine, one for each of its paths. Each row in the letter consists of a \emph{block number} in $\{0,\ldots,p\}$, two \emph{register values} (one for the read and one for the write operations), a \emph{successor state} and an \emph{index of a row} in an associated trace word (word in $\galph^*$).
Let $\widetilde \Sigma =  \galph \times \calph$ be the product of the trace and execution alphabets.

In what follows, if $\widetilde \tau = \widetilde \sigma_1\ldots\widetilde\sigma_f \in \galph^*$, then 
$\widetilde \sigma_j = (\widetilde \sigma_{1,j},\ldots,\widetilde \sigma_{t,j})$ 
denotes the elements of the $j$-th letter of the word $\widetilde \tau$ for $j \in \{1,\ldots,f\}$, and we use
$\widetilde \tau_i  = \widetilde \sigma_{i,1}\ldots\widetilde\sigma_{i,f}$ 
to denote the $i$-th row of $\widetilde\tau$ for $i \in \{1,\ldots,t\}$.
Similarly, if $\widetilde \tau = \widetilde \sigma_1\ldots\widetilde\sigma_f \in {\widetilde \Sigma}^*$, 
the $j$-th letter is
$\widetilde \sigma_j = (\widetilde \sigma_{1,j},\ldots,\widetilde\sigma_{t,j},
\widetilde\eta_{1,1,j},\ldots,\widetilde\eta_{1,\widetilde r,j},\ldots,
\widetilde\eta_{m,1,j},\ldots,\widetilde\eta_{m,\widetilde r,j}),$
for $j \in \{1,\ldots,f\}$, 
and the $i$-th row is
$\widetilde \tau_i  = \widetilde \sigma_{i,1}\ldots\widetilde\sigma_{i,f}$,
for $i \in \{1,\ldots,t\}$.
For $n \in \{1,\ldots, m\}$, $h \in \{1,\ldots,\widetilde r\}$ and $j\in \{1,\ldots,f\}$,
the corresponding letter from $\calph$ is denoted
$\widetilde\eta_{n,h,j} = (p_{n,h,j},b_{n,h,j},b'_{n,h,j},q'_{n,h,j},t_{n,h,j})$.

In the remainder of this section we present the intuition behind the automata 
constructions and their properties.

\subsection{PDA accepting traces in a graph}
The PDA $\mc P_{\mathsf{t}}$ is the intersection of a PDA $\mc P$ obtained from ${\mc G}^k$, where
$\mc G = (V,\Sigma,R,G_0)$ is an SPGG, and a NFA $\mc A_{\mathsf r}$ that checks that register identifiers are correctly placed.

\looseness=-1
The construction of $\mc P = (Q_p,q^0_p,\galph\ \dot\cup \ \{\vdash,\dashv\},\galph\ \dot\cup \ \{\vdash,\dashv\} \ \dot\cup \ \widetilde V \ \dot\cup \ \{\bot\},\bot,\delta_p)$ resembles the classical construction of a PDA given a CFG. Here, instead of words generated by a CFG the language $\mc L(\mc P)$ of $\mc P$ consists of $t$-tuples of (annotated) traces in some graph generated by the grammar. 
The automaton has a stack alphabet $\galph\ \dot\cup \ \{\vdash,\dashv\} \ \dot\cup \ \widetilde V \ \dot\cup \ \{\bot\}$, where $\widetilde V$ consists of symbols corresponding to the variables in $\mc G$. 
The transitions in $\delta_p$ can be grouped according to the top symbol on the stack: empty stack, top symbol $\widetilde \sigma \in \galph \cup \{\vdash,\dashv\}$, and top symbol $\widetilde v \in \widetilde V$. 
Transitions for $\widetilde v \in \widetilde V$ correspond to the production rules of the SPGG $\mc G$.
For the series composition $\delta_p$ employs the additional symbol $\flat$ to allow for traces that are aligned in a way that letters in $\{1,\ldots,t\}^t$ reflect the information about nodes shared by the respective paths.
For the parallel composition $\delta_p$ guesses symbols, in the graphs generated by which the corresponding traces occur, together with the number of traces in the subgraph generated by each symbol. The number of times a new branch is introduced is bounded by $t$, the number of parallel traces.  

\looseness=-1
$\mc P$ does not check that the register identifiers in the annotation are consistent among letters corresponding to edges in the graph that share a node, i.e., that letters corresponding to these edges have the same identifier for this node. This is done by the NFA  
$\mc A_{\mathsf r} = (Q_{\mathsf r},q^0_{\mathsf r},\galph\ \dot\cup \ \{\vdash,\dashv\},\delta_{\mathsf r},F_{\mathsf r})$, 
which also verifies that identifiers for different nodes are unique. 
To this end, each state $\widetilde q$ of $\mc A_{\mathsf r}$ contains a path index $l_h \in \{1,\ldots,t\}$ and a register identifier $i_h \in \{0,\ldots,k\}$ for each row $\widetilde\tau_h$ of $\widetilde\tau$. $\delta_{\mathsf r}$ checks that 
the letters in $\widetilde \tau$ that correspond to edges incident with the same node agree on the corresponding register identifier. The path indices $l_h$ in $\widetilde q$ are used to identify branching or joining paths and the register identifiers $i_h$  to check the required equalities. In the accepting states the equalities for the sink node of the graph must be satisfied. 
Additionally, $\delta_{\mathsf r}$ verifies that the register identifiers in $\widetilde \tau$ corresponding to different nodes are different. 

\looseness=-1
The PDA $\mc P_{\mathsf{t}}$ has $\mc L(\mc P_{\mathsf{t}}) = \mc L(\mc P) \cap \mc L(\mc A_{\mathsf r})$.
The construction of $\mc P$ and $\mc A_{\mathsf r}$ ensures that if $\widetilde\tau \in \mc L(\mc P_{\mathsf{t}})$, then
there exists $(G,\kappa) \in \mc L^u({\mc G}^k)$ and 
for each $i \in \{1,\ldots,t\}$ there exists a sequence of nodes $\widetilde \pi_i$ in $G$ such that
for each row $\widetilde \tau_i$ of $\widetilde\tau$ there exists a subsequence $\pi_i \in \paths(G,\tau_i)$ of $\widetilde \pi_i$ corresponding to the projection 
$\tau_i = \big(\widetilde\tau_i|_{\Sigma \times \{0,\ldots,k\}^2\times \{1,\ldots,t\}} \big) |_{\Sigma}$ of $\widetilde \tau_i$ on $\Sigma$. 
Furthermore, these paths can be chosen such that edges corresponding to rows with the same path index connect the same pair of nodes. 
Additionally, the mapping $\kappa$ for the nodes on these paths agrees with the corresponding register identifiers in $\widetilde \tau$.

Conversely, if $(G,\kappa) \in \mc L^u(\mc G^k)$ and 
for every $n \in \{1,\ldots,m\}$ and $h \in \{1,\ldots,\widetilde r\}$ we are given a path $\widehat\pi_{n,h} \in \paths(G,\widehat\tau_{n,h})$ for some trace $\widehat\tau_{n,h} \in \Sigma^*$, 
then there exists a word $\widetilde\tau \in \mc L(\mc P_{\mathsf{t}})$, which corresponds to these paths and traces.
The word $\widetilde \tau$ is obtained by ordering, extending  and annotating the given traces.
\label{sec:pda-traces}

\subsection{2NFA accepting executions}
We construct a 2NFA $\widetilde{\mc A_n}$ for each $n \in \{1,\ldots,m\}$ that checks that the sequence described by the rows of the word that correspond to $n$ is indeed an execution of $\mc M$ that reads the corresponding rows of the trace word.
Furthermore, $\widetilde{\mc A_n}$ verifies that the machine  switches between traces described by different rows of the trace word only at positions at which the traces share a node in the corresponding paths. 

Each state $\widetilde q$ of 
$\widetilde {\mc A_n} = (\widetilde Q_n,\widetilde Q^0_n,\widetilde \Sigma\ \dot\cup \ \{\vdash,\dashv\},\widetilde \delta_n,\widetilde Q_n)$
contains a state $q \in Q$ of $\mc M$, which is the current state of the simulated machine, and an index $i \in \{1,\ldots,t\}$ in the trace-word that is part of the input word.
$\widetilde \delta_n$ refers to the transition relation $\delta$ of $\mc M$ to check the existence of a transition of $\mc M$ that performs the read and write operations determined by the read letter of $\widetilde\tau$.
The state $q$ is updated according to $\delta$ and remains unchanged when the machine is inactive in the current part of the trace. The row of the trace word that is read in state $\widetilde q$ is determined by $i$. The index $i$ can be changed by $\widetilde \delta_n$ only if for the current letter we have $\widetilde \sigma_{t_{n,h}} \in \{1,\ldots,t\}$ and $p_{n,h} > 0$, and for the new value $i'$ it must hold that $\sigma_{i'} = \sigma_{i}$. That is, the machine can switch between traces only at positions where the paths intersect. 
An additional component of $\widetilde q$ is used to check that block number $0$ in $\widetilde \tau$ is used to correctly encode the reversals of the machine (which do not have to be at the start or sink nodes of the graph). All states are accepting.

\looseness=-1
Then,  $\widetilde \tau \in \mc L(\widetilde {\mc A_n})$ iff by taking the
elements of $\widetilde \tau$ corresponding to machine $n$ in the appropriate order
we can construct an execution $\xi_n$, formally defined as follows.
For each $h \in \{1,\ldots,\widetilde r\}$ and $l \in \{1,\ldots,f\}$, 
if $\widetilde \sigma_{t_{n,h,l},l}  = (\sigma,c,j_1,j_2) \in \Sigma \times \{1,\ldots,t\} \times \{0,\ldots,k\}^2$ and $p_{n,h,l}  > 0$, then,
if $h$ is odd, then 
$\widehat \xi_{n,h,l} = (\sigma,b_{n,h,l},b'_{n,h,l}) \cdot q'_{n,h,l}$,
and if $h$ is even, then 
$\widehat \xi_{n,h,l} = q'_{n,h,l} \cdot (\sigma,b_{n,h,l},b'_{n,h,l})$,
Otherwise, $\widehat \xi_{n,h,l} = \varepsilon$.
Then
$\widehat\xi_{n,h} = \widehat\xi_{n,h,1}\cdot\ldots\cdot\widehat\xi_{n,h,f}$ if $h$ is odd, and
$\widehat\xi_{n,h} = \widehat\xi_{n,h,f}\cdot\ldots\cdot\widehat\xi_{n,h,1}$ otherwise.
Finally, 
$\xi_n = q^0\cdot\widehat\xi_{n,1}\cdot\ldots\cdot\widehat\xi_{n,\widetilde r}$.
\label{sec:2nfa-executions}

\subsection{2NFA accepting valid read-write sequences}
Here we describe a 2NFA $\widetilde{\mc A_{\mathsf c}}$ that checks that the executions of the different machines described by the input word are compatible with each other. That, is that the read and write operations of different machines match when executed in the order determined by the input word, where each operation is labelled with a block number. $\widetilde{\mc A_{\mathsf c}}$ verifies that each block number is used in a single execution and that for each execution the sequence of positive block numbers is nondecreasing. To check the validity of the corresponding read-write sequence w.r.t.\ the initial register values, $\widetilde{\mc A_{\mathsf c}}$ tracks the register values at the end and at the beginning of each block and compares the values at the beginning of block $i+1$ with those at the end of block $i$. 
An \emph{assumption} is a partial function $A : \{1,\dots,p\} \to \mathbb{B}^k$ that maps a block number to a valuation of the registers, representing the obligation to verify that at the beginning of a block the registers have the respective values. Similarly, a \emph{guarantee} is a function $G : \{1,\dots,p\} \to \mathbb{B}^k$ used to propagate the guarantee that at the end of a block the registers have a certain value.

Each state of the automaton $\widetilde{\mc A_{\mathsf c}} = (\widetilde Q_{\mathsf c},\widetilde{q^0_{\mathsf c}},\widetilde \Sigma\ \dot\cup \ \{\vdash,\dashv\},\widetilde{\delta_{\mathsf c}},\widetilde{F_{\mathsf c}})$ contains 
a block number $p_n$ and a valuation of the registers $\beta_n$ for machine $n$, 
a set $P$ of already seen block numbers,
an assumption $A$ and a guarantee $G$.
The transition relation $\widetilde{\delta_{\mathsf c}}$ checks that all read operations of machine $n$ except those at the beginning of a block read the value stored in $\beta_n$. 
At the beginning of a block of machine $n$, $\widetilde{\delta_{\mathsf c}}$ guesses a valuation of the registers for read operations and stores them in $\beta_n$. The new block number and the guess are added to the set $A$. 
The values of its write operations are used to update $\beta_n$ and, at the end of a block the respective guarantee is added to $G$. $\widetilde{\delta_{\mathsf c}}$ discharges assumptions in $A$ for which the respective guarantees are in $G$. 
In an accepting state the set $A$ should be empty and 
the set $P$ of all block numbers in $\widetilde \tau$ should contain all block numbers smaller or equal the maximal one.

By construction, in each word $\widetilde \tau  \in \mc L(\widetilde{\mc A_{\mathsf c}})$ each block number is assigned to at most one machine and for each machine the sequence of positive block numbers is nondecreasing. All such words $\widetilde\tau$ are accepted by $\widetilde{\mc A_{\mathsf c}}$ iff the read-write sequence, constructed by ordering elements of $\widetilde \tau$ according to block number while preserving the order for each individual machine, is valid w.r.t. the initial register contents.\label{sec:2nfa-read-write}

\subsection{NFA checking configuration properties}
The NFA $\mc A_{\mathsf s} = (\widetilde Q_{\mathsf s},\widetilde{q^0_{\mathsf s}},\widetilde \Sigma\ \dot\cup \ \{\vdash,\dashv\},\widetilde{\delta_{\mathsf s}},\widetilde{F_{\mathsf s}})$ checks that in some run in $T(\mc M,m,{\mc G}^k)$ corresponding to the input word, a configuration that satisfies the given configuration property $F = \exists n.\ S \preceq \mu(n)$ is reached.

Since the configuration property $F$ asserts the existence of a node in the graph of a configuration, 
potential such configurations can be detected by inspecting (at most) two consecutive letters in the word.
The information relevant for the satisfaction of a configuration property consists
of the block number and successor state components of 
the letters of the execution word and the letters of the trace word. Thus, we define the set 
$C = \{1,\ldots,p\}^t \times (Q \cup \{\bot\})^t \times \galph$ and consider pairs of elements of $C$.

\looseness=-1
Let  
$c_0 = (p_{1,1}^0,\ldots,p_{m,\widetilde r}^0,
q_{1,1}^0,\ldots,q_{m,\widetilde r}^0,
\widetilde \sigma_{1,1}^0,\ldots,\widetilde \sigma_{m,\widetilde r}^0)$, 
$c_\bot = (p_{1,1}^\bot,\ldots,p_{m,\widetilde r}^\bot,
q_{1,1}^\bot,\ldots,q_{m,\widetilde r}^\bot,
\widetilde \sigma_{1,1}^\bot,\ldots,\widetilde \sigma_{m,\widetilde r}^\bot)$, 
where for $n \in \{1,\ldots,m\}$ and $h \in \{1,\ldots,\widetilde r\}$,
$p_{n,h}^0 = p_{n,h}^\bot = 0$,
$q_{n,h}^0 = q^0$, $q_{n,h}^\bot = \bot$,
$\widetilde \sigma_{n,h}^0 = \widetilde \sigma_{n,h}^\bot = (\flat,1,0,0)$.

Let us consider two elements of the set $C$:
$c'  =  (p_{1,1}',\ldots,p_{m,\widetilde r}',q_{1,1}',\ldots,q_{m,\widetilde r}',
\widetilde \sigma_{1,1}',\ldots,\widetilde \sigma_{m,\widetilde r}') \in C$ and
$c'' =  (p_{1,1}'',\ldots,p_{m,\widetilde r}'',q_{1,1}'',\ldots,q_{m,\widetilde r}'',
\widetilde \sigma_{1,1}'',\ldots,\widetilde \sigma_{m,\widetilde r}'') \in C$.

We say that the pair $(c',c'')$ \emph{occurs in 
$\widetilde\tau = \widetilde{\sigma_1}\ldots\widetilde{\sigma_f} \in \widetilde \Sigma$} iff
there exists a sequence of consecutive letters in $\widetilde\tau$ such that those of $c'$ and $c''$ that are not 
equal to $c_0$ and $c_\bot$ match these letters of $\widetilde\tau$ in the same order. 
Formally, $(c',c'')$ \emph{occurs in $\widetilde \tau$} iff one of the following conditions is satisfied.
\begin{compactitem}
\item[(1)] $c' = c_0$, and 
$p_{n,h}'' = p_{n,h,1}$, 
$q_{n,h}'' = q_{n,h,1}$ and 
$\widetilde \sigma_{n,h}'' = \widetilde\sigma_{t_{n,h,1},1}$ ($c''$ matches $\widetilde \sigma_1$).
\item[(2)] $c'' = c_\bot$, and
$p'_{n,h} = p_{n,h,1}$, 
$q'_{n,h} = q_{n,h,1}$ and 
$\widetilde \sigma'_{n,h} = \widetilde\sigma_{t_{n,h,1},f}$ ($c'$ matches $\widetilde \sigma_f$).
\item[(3)] There exists $1 < l \leq f$ such that 
\begin{compactitem}
\item $p'_{n,h} = p_{n,h,l-1}$, 
$q'_{n,h} = q_{n,h,l-1}$ and
$\widetilde \sigma'_{n,h} = \widetilde\sigma_{t_{n,h,l-1},l-1}$ ($c'$ matches $\widetilde \sigma_{l-1}$),
\item $p''_{n,h} = p_{n,h,l}$, 
$q''_{n,h} = q_{n,h,l}$ and
$\widetilde \sigma''_{n,h} = \widetilde\sigma_{t_{n,h,l},l}$ ($c''$ matches $\widetilde \sigma_l$).
\end{compactitem}
\end{compactitem}

\looseness=-1
Consider a configuration $\gamma \in \Gamma$ of a run $\rho$ that satisfies the configuration property $F$. 
This means that there exists an edge $e \in E$, such that some of the nodes $\src(e)$ and $\trg(e)$
makes the property true. Furthermore, there exits a set of machines involved in the satisfaction of the
property in $\gamma$. Among these machines, we distinguish between the one that executed the last 
transition in $\rho$ leading to this configuration and the remaining machines. By the definition of
runs of $T(\mc M,m,{\mc G}^k)$, the current node and states of these remaining machines should be
reached at the end of one of their execution blocks. We define a predicate about pairs of 
elements of $C$, sets of machines and corresponding positions in their executions (i.e., rows in the 
respective letter of the execution word). The automaton $\mc A_s$ will use this predicate to 
identify letters of the word that may encode configurations satisfying the configuration property F.

Let $S \in \mathbb{M}(Q)$, $M \subseteq \{1,\ldots,m\}$ and $f_M : M \to \{1,\ldots,\widetilde r\}$.
For each $n \in M$, let $f_n = f_M(n)$ and 
if $f_n$ is odd, then $p_n = p_{n,f_n}'$, $q_n = q_{n,f_n}'$ and $\sigma_n = \widetilde \sigma_{n,f_n}'$, 
and if $f_n$ is even, then $p_n = p_{n,f_n}''$, $q_n = q_{n,f_n}''$ and $\sigma_n = \widetilde \sigma_{n,f_n}''$.
Let $n_0 \in M$ be such that for each $n \in M$, it holds that $p_n \leq p_{n_0}$. 
We define $p_n(c',c'',M,f_M) = p_n$ for each $n \in M$ and $n_0(c',c'',M,f_M) = n_0$.

The \emph{node predicate} $\nprop(S,c',c'',M,f_M)$ is true iff the conditions listed below hold.
\begin{compactitem}
\item $S = [q_n \mid n \in M]$ and 
for each $n \in M \setminus \{n_0\}$, $p_n < p_{n_0}$ and $p'_{n,f_n} \neq p''_{n,f_n}$.
\item One of the following requirements is satisfied:
\begin{compactitem}
\item $\sigma'_{n_0} \in \{1,\ldots,t\}$ and $\sigma_n'  = \sigma'_{n_0}$ for each $n \in M$.
\item $\sigma'_{n_0} = (\sigma_0,l_0,j_0,j'_0) \in \Sigma \times \{1,\ldots,t\} \times \{0,\ldots,k\}^2$ and for each $n \in M$ there exist $\sigma\in\Sigma,j,j' \in \{0,\ldots,k\}$ such that  $\sigma_n'  = (\sigma,l_0,j,j')$.
\item $\sigma''_{n_0} \in \{1,\ldots,t\}$ and $\sigma_n''  = \sigma''_{n_0}$ for each $n \in M$.
\item $\sigma''_{n_0} = (\sigma_0,l_0,j_0,j'_0)\in \Sigma \times \{1,\ldots,t\} \times \{0,\ldots,k\}^2 $ and for each $n \in M$ there exist $\sigma\in\Sigma,j,j' \in \{0,\ldots,k\}$ such that  $\sigma_n''  = (\sigma,l_0,j,j')$.
\end{compactitem}
\end{compactitem}

\smallskip

In order to evaluate the above predicates on positions of the input word $\widetilde \tau$,
the NFA $\mc A_{\mathsf s}$ stores in its state an element $c'$ of $C$. 
The current letter of the input word determines an element $c''$ of $C$, 
and $\mc A_{\mathsf s}$ evaluates the node predicate on the pair $(c',c'')$. 
In order to verify that some run corresponding to $\widetilde\tau$ contains a 
configuration satisfying the configuration property $F$, $\mc A_{\mathsf s}$ 
must detect all pairs on which the predicate $\nprop$ holds true, that is, they
might encode a final configuration. For each such pair $\mc A_{\mathsf s}$ must  
verify that some of the pairs on which the predicate $\nprop$ holds actually fulfils 
a global condition on $\widetilde \tau$. Namely, that for all the involved machines, 
the currently executed block number is the last one. (We can, w.l.o.g.\,
restrict to runs where only the last configuration can be final.)

A state $\widetilde q$ of $\mc A_{\mathsf s}$ contains an element $c'$ of $C$, 
a boolean value $final \in \mathbb B$, and for each machine $n$ it contains components 
$P_n \in 2^{\{1,\ldots,p\}}$ and 
$p_n \in \{0,\ldots,p\}$.
The set $P_n$ consists of the already seen block numbers for $n$. 
If the current letter defines $c \in C$ such that $\nprop(S,c',c,M,f_M)$ holds 
true for some $M \subseteq \{1,\ldots,m\}$ and some function $f_M$, then 
$final$ can be set to $1$ if it is $0$, and the current block number of machine $n$ 
for each $n \in M$ can be stored in $p_n$, in order to verify later that a final
configuration is indeed reached (by checking that $p_n$ is the maximal block number for $n$).
Based on the currently read letter of $\widetilde \tau$, the transition relation 
$\widetilde{\delta_{\mathsf s}}$ updates the component $c'$ of the state.
The accepting state $\widetilde q_{\mathsf s}^f$ can be entered after reading $\dashv$
if  $final = 1$ and if $p_n$ is the maximal block number for machine 
$n$ for each $n$.

The construction of $\mc A_{\mathsf s}$ ensures that $\widetilde \tau \in \mc L(\mc A_{\mathsf s})$ iff
there exists a pair $(c',c'')$ occurring in $\widetilde \tau$ encoding a configuration
that satisfies the given configuration property $F$.
\label{sec:1nfa-properties}

\subsection{Correctness of the algorithm}
\looseness=-1
Let $\mc A_1,\ldots,\mc A_m$ be an NFA obtained respectively from $\widetilde{\mc A}_1,\ldots,\widetilde{\mc A}_m$ such that for each $i \in \{1,\ldots,m\}$, $\mc L(\mc A_i)  = \mc L(\widetilde{\mc A}_i)$. Let $\mc A_{\mathsf c}$ be an NFA constructed from $\widetilde{\mc A}_{\mathsf c}$ such that $\mc L(\mc A_{\mathsf c})  = \mc L(\widetilde{\mc A}_{\mathsf c})$. We then construct the NFA $\mc A_{\mathsf e}$ by intersecting $\mc A_1,\ldots,\mc A_m$, $\mc A_{\mathsf c}$ and $\mc A_{\mathsf s}$ and projecting the result on $\galph$, i.e., $\mc L(\mc A_{\mathsf e}) = \big(\bigcap_{i=1}^{m} \mc L(\mc A_i) \cap \mc L(\mc A_{\mathsf c}) \cap L(\mc A_{\mathsf s}) \big)|_{\galph}$. 
The PDA $\mc A$ is the intersection of $\mc P_{\mathsf t}$ and $\mc A_{\mathsf e}$.

\begin{theorem}\label{thm:correctness}
$\mc L(\mc A) \neq \emptyset$ iff 
there exists an $r$-reversal bounded run $\rho = \gamma_0 \ldots \gamma_f$ in  $T({\mc M}, m, {\mc G}^k)$ 
such that $\gamma_i \models F$ for some $0 \leq i \leq f$, i.e., a run that reaches a configuration satisfying $F$.
\end{theorem}

\section{Conclusion.}
In this paper we define and study a class of concurrent finite-state automata
traversing series-parallel graphs and communicating through shared finite registers 
located at the nodes of the graph.
We considered a model in which a \emph{fixed number} of finite-state 
machines traverse the nodes of a \emph{series-parallel graph}. 
The series-parallel graphs are generated by a graph grammar,
and as we do not impose an a priori bound on the size of the graphs,
the resulting system is infinite-state. 
Since the emptiness problem for this model is in general undecidable,
we consider a natural restriction by putting bounds on the number of reversals
along the computation and the number of shared registers in the graph.
With these two restrictions, we show that the emptiness problem is decidable and
can be reduced to PDA emptiness.

As we noted in Section~\ref{sec:bounded-verification}, our decidability result
holds for a more general model of communication between the machines, in which
either read or write (but not both) operations on registers can be non-local, 
that is, access a register that is not at the node where the machine is currently located.
Another possible extension that we omitted for simplicity concerns the language of configuration
properties. While here we consider properties that quantify over individual nodes in the graph,
we can, in principle, extend the construction described in Section~\ref{sec:1nfa-properties} to
handle configuration properties asserting the existence of edges with certain labels, or a fixed number
of adjacent nodes and edges.

Interesting directions for future work include establishing the complexity of the bounded emptiness problem
for our model, as well as studying different extensions. One possibility is to allow 
parametrization in the number of concurrent machines, another is to consider other classes of
context-free GTSs. For example, using the techniques from~\cite{MadhusudanP11/auxiliarystorage} one can try
to extend our results to a more general class of graphs of bounded tree width.

\bigskip
{\bf Acknowledgements.} We thank the anonymous reviewers for their helpful and insightful comments.

\bibliographystyle{eptcs}
\bibliography{main.bib}

\end{document}